\documentclass[12pt]{iopart}

\usepackage{iopams}  
\usepackage[dvipdfmx]{graphicx}
\begin{document}

\title[Film figure of merit]{Thickness optimization of the output power and effective thermoelectric figure of merit of thin thermoelectric generator}

\author{Kazuhiko Seki}

\address{GZR, National Institute of Advanced Industrial Science and Technology (AIST), Onogawa 16-1 AIST West, Ibaraki, 305-8569, Japan}
\ead{k-seki@aist.go.jp}

\author{Masakazu Mukaida}
\address{NMRI, National Institute of Advanced Industrial Science and Technology (AIST), Higashi 1-1-1 AIST Central 5, 305-8565, Japan}

\author{Qingshuo Wei}
\address{NMRI, National Institute of Advanced Industrial Science and Technology (AIST), Higashi 1-1-1 AIST Central 5, Ibaraki, 305-8565, Japan}

\author{Takao Ishida}

\address{GZR, National Institute of Advanced Industrial Science and Technology (AIST), Onogawa 16-1 AIST West, Ibaraki, 305-8569, Japan}

\vspace{10pt}
\begin{indented}
\item[]February 2022
\end{indented}

\begin{abstract}
The conventional thermoelectric figure of merit and the power factor are not sufficient as a measure of thin film quality of thermoelectric materials, where the power conversion efficiency depends on the film dimensions. By considering the film size, the effective thermoelectric figure of merit and effective Seebeck coefficient are introduced to guarantee that the maximum energy conversion efficiency increases as the effective thermoelectric figure of merit increases. Similarly, the effective power factor is defined. By introducing typical material properties for Bi$_2$Te$_3$ and PEDOT, we study the thickness dependence of the effective figure of merit and the effective power factor. 
\end{abstract}

%
%
%
%
%

Thermoelectric generators directly convert waste thermal energy into electrical power. 
Electric currents can be generated from temperature gradients by thermoelectric effect;  
the energy conversion efficiency is known to be characterized by the dimensionless thermoelectric figure of merit, 
given by 
\cite{Ioffe,Subramanian_06}
\begin{eqnarray}
\mbox{zT}_{\rm b}=\frac{\alpha^2 T_{\rm m}}{\rho \kappa}, 
\label{eq:zTbulk}
\end{eqnarray}
where $T_{\rm m}$ is the mean temperature between the cold side and the hot side, $\alpha$ is the Seebeck coefficient, $\rho$ is the electrical resistivity, 
and $\kappa$ is the thermal conductivity. 
The thermoelectric figure of merit $\mbox{zT}_{\rm b}$ provides a measure of the quality of thermoelectric materials and 
can be used for the guide to compare thermoelectric properties of materials. 

The recent demand for thermoelectric generators for Internet of Things (IoT) and wearable devices 
has urged to develop 
thin film thermoelectric devices, which are flexible and could operate around room temperature. \cite{Goncalves_10,Ito_17,YAMASHITA_11,Sungtaek_00,Mukaida_20,Wang_19,Burton_22}
Such devices should work under limited temperature gradients and require sufficient energy conversion efficiency. 
Conventionally, 
the thermoelectric figure of merit 
and the power factor have been used to 
relate the material properties such as the electrical resistivity and the thermal conductivity   
with the maximum energy conversion efficiency and the maximum output power, respectively.  
However, for thin film thermoelectric devices, 
the electrical contact resistivity and the heat transfer on the surface facing to the environments need to be considered because of large areal size compared to the film thickness. \cite{Ito_17,Gurevich_12,Sungtaek_00,Mukaida_20}
Therefore, the energy conversion efficiency could be influenced by the film dimensions 
in addition to the thermoelectric figure of merit $\mbox{zT}_{\rm b}$ of the bulk material. 
To consider the influence of film dimensions on the energy conversion efficiency, 
we define below 
$\mbox{zT}_{\rm eff}(d)$, 
where $d$ indicates the film thickness. 

We first introduce the effective electrical resistivity using 
the intrinsic material electrical resistivity  
$\rho$ $\Omega $m 
and the contact electrical resistivity  
$\rho_{\rm c}$ $\Omega \mbox{m}^2$. 
Since they are connected electrically in series, 
the effective resistivity can be expressed as 
$\rho_{\rm eff}=\rho+\rho_{\rm c}/d$. 
When the electrical contact resistivity of one side is given by 
$\rho_{{\rm c}0}$ and that of the other side is given by $\rho_{{\rm c}d}$, 
the total electrical contact resistivity is expressed as
$\rho_{\rm c}=\rho_{{\rm c}0}+\rho_{{\rm c}d}$. 
Similarly, we define the effective thermal conductivity 
$\kappa_{\rm eff}$ 
by the intrinsic material thermal conductivity  
$\kappa$ W/(mK) and 
the heat transfer coefficient $h$ W/(m$^2$K) as 
 $\kappa_{\rm eff}=(1/\kappa+1/(hd))^{-1}$, 
 where the heat transfer coefficient is inverse of the contact thermal resistivity;  
 the effective thermal conductivity is given by the inverse of the sum of the intrinsic material thermal resistivity and the contact thermal resistivity.
  When the heat transfer coefficient of one side is given by 
$h_{0}$ and that of the other side is given by $h_{d}$, 
the total heat transfer coefficient is expressed as
$h=\left(1/h_{0}+1/h_{d}\right)^{-1}$. 
Previously, the effective thermal conductivity was given by the sum of the intrinsic material thermal conductivity and the thermal interface conductance, 
 which indicates parallel thermal flows of these components. 
 \cite{YAMASHITA_11}

\begin{figure}[h]
\includegraphics[width=12cm]{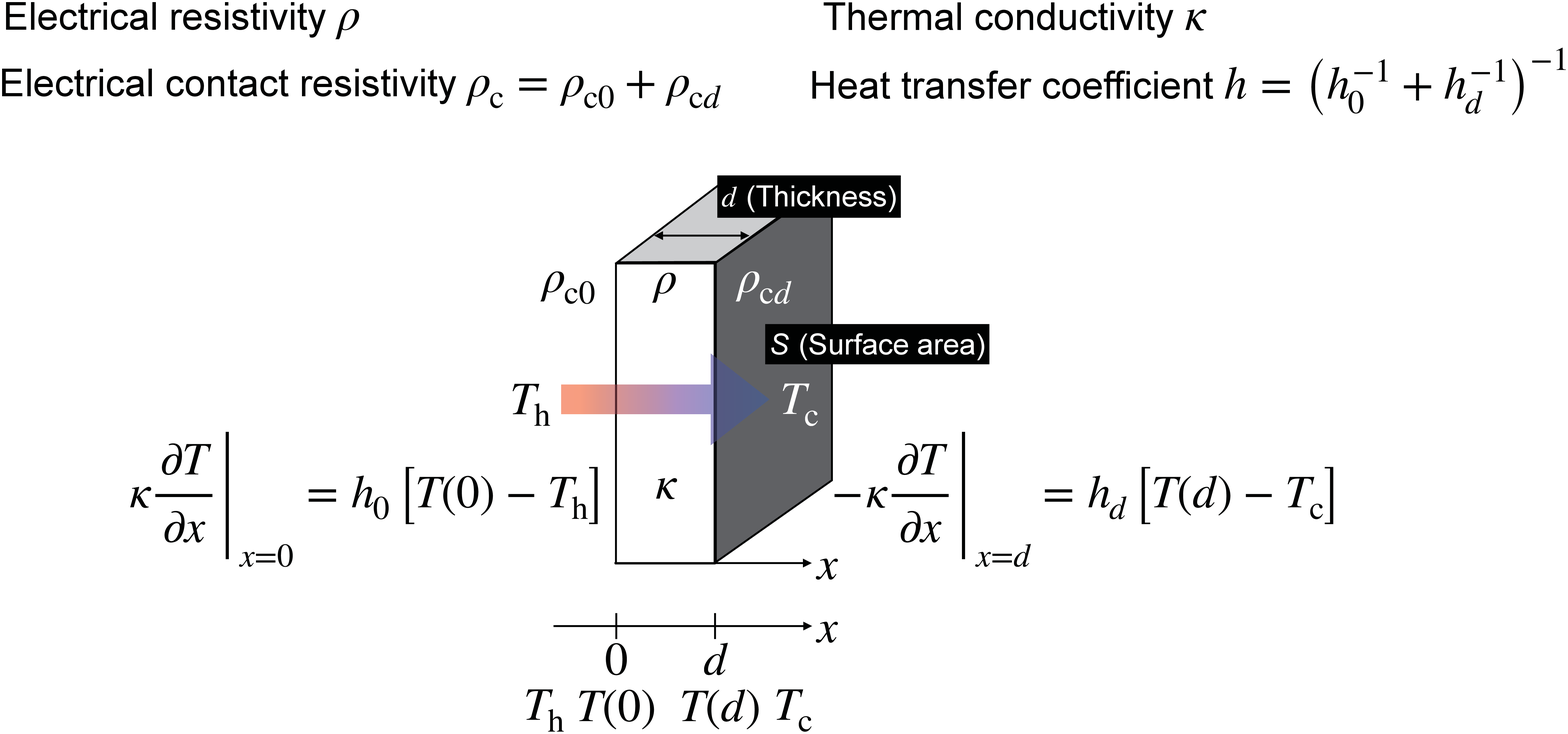}
\caption{(Color online) 
A slab of thermoelectric material (thickness $d$ and cross section area $S$) under thermal gradient. 
The environmental temperature facing at $x=0$ is denoted by $T_{\rm h}$ and that facing at $x=0$ is denoted by $T_{\rm c}$.
Heat flows under the temperature gradient, where the thermal conductivity and the heat transfer coefficient are denoted by 
$\kappa$ and $h$, respectively.
Charges flow under the electromotive force generated by the temperature difference, where the electrical resistivity and 
the electrical contact resistivity are denoted by $\rho$ and $\rho_{c}$, respectively. 
}
\label{fig:ZT_slab}
\end{figure}
We introduce the figure of merit by considering the maximum of energy conversion efficiency by taking into account the film dimensions. 
For thick bulk  materials, the maximum conversion efficiency is known to be an increasing function of 
zT$_{\rm b}$ at the mean temperature of the cold and the hot side. 
Therefore, the maximum conversion efficiency is higher as zT$_{\rm b}$ increases. 
We needs to define zT$_{\rm eff}(d)$  for films in such a way 
that the maximum conversion efficiency increases by increasing zT$_{\rm eff}(d)$. 

We consider the case that the film of thickness $d$ and the surface area of $S$ is located between $x=0$ and $d$ in the x coordinate 
as shown in Fig. \ref{fig:ZT_slab}.
The ambient temperature around $x=0$ is $T_{\rm h}$ and the ambient temperature around $x=d$ is 
$T_{\rm c}$; we assume $T_{\rm h}>T_{\rm c}$. 
We also assume that systems operate under a constant temperature difference rather than under constant heat flux. \cite{Gao_22}
The energy conversion efficiency can be expressed as \cite{Sherman_66}
\begin{eqnarray}
\eta=\frac{V I}{\alpha T(0) I-RI^2/2+K\left[ T(0) - T(d)
\right]}, 
\label{eq:efficiency1}
\end{eqnarray}
where $V$ and $I$ denote the voltage across the film and the electrical current passing through the film, respectively; 
$R$ and $K$ denote the electrical resistance and the thermal conductance, respectively. 
The temperature at $x=0$ and $x=d$ is denoted by $T(0)$ and $T(d)$, respectively;  
$T(d)$ differs from the environmental temperature denoted by $T_{\rm c}$ if the heat transfer coefficient is low. 
We calculate $T(0)$ and $T(d)$ when the heat transfer coefficient at the hot side ($x=0$) is $h_{\rm h}$ and 
that at the cold side ($x=d$) is $h_{\rm c}$. 
$T(0)$ reduces to $T_{\rm h}$ in the limit of $h_{\rm h} \rightarrow \infty$. 
As shown in \ref{Sec:ApA}, we obtain 
\begin{eqnarray}
T(0)-T(d)=\Delta T/(1+\kappa/(h d)), 
\label{eq:dT}
\end{eqnarray}
where $\Delta T=T_{\rm h}-T_{\rm c}$ and 
$h=\left(1/h_0+1/h_d\right)^{-1}$. 
We also find 
\begin{eqnarray}
T_{\rm m}=(T(0)+T(d))/2=(T_{\rm h}+T_{\rm c})/2. 
\label{eq:meanT}
\end{eqnarray}

When the thermoelectric power is generated, the external load with the resistance $R_{\rm ext}$ is imposed. 
We introduce $R_{\rm ext}= y R_{\rm eff}$, where $R_{\rm eff}=(\rho +\rho_{\rm c}/d)d/S$ 
for the film with the thickness $d$ and the surface area of $S$ 
facing to the environments. \cite{Ioffe,Angrist_76,Kim_15,Terasaki}
We have $I=\alpha [T(0)-T(d)]/[(1+y)R_{\rm eff}]$ and $V=\alpha [T(0)-T(d)]-I R_{\rm eff}
=\alpha [T(0)-T(d)]y/(1+y)$. 
The power can be expressed as 
\begin{eqnarray}
P=IV=\frac{(\alpha_{\rm eff} \Delta T)^2}{R_{\rm eff}} \frac{y}{(1+y)^2}, 
\label{eq:power}
\end{eqnarray}
where $\alpha_{\rm eff}$ is defined by 
\begin{eqnarray}
\alpha_{\rm eff}=\frac{\alpha}{1+\kappa/(h d)} .
\label{eq:alphaeff}
\end{eqnarray}
By introducing $\alpha_{\rm eff}$,  the thermoelectric power can be expressed in the conventional form as shown in Eq. (\ref{eq:power}). 

Using $y=R_{\rm ext}/R_{\rm eff}$, 
the energy conversion efficiency given by (\ref{eq:efficiency1}) can be expressed for films as
\begin{eqnarray}
\eta(y)=\frac{y\Delta T}{(1+y)T_{\rm m} +(1+y)^2/Z_{\rm eff} + y \Delta T/2},  
\label{eq:efficiencyfilm}
\end{eqnarray}
where $Z_{\rm eff}$ is defined by  
\begin{eqnarray}
Z_{\rm eff} =\frac{\alpha_{\rm eff}^2}{R_{\rm eff} K_{\rm eff}}, 
\label{eq:Zeff}
\end{eqnarray}
and we have $R_{\rm eff}=(\rho +\rho_{\rm c}/d)d/S$ and $K_{\rm eff}=\left(\kappa^{-1}+1/(h d)\right)^{-1}S/ d$. 
Using $d \eta(y)/dy=0$, \cite{heikes_61} 
we find $y$ at the maximum conversion efficiency as 
$y_{\rm max}=\sqrt{1+Z_{\rm eff} T_{\rm m}}$; the maximum conversion efficiency is obtained as 
\begin{eqnarray}
\eta_{\rm max}&=
\frac{\Delta T \sqrt{1+Z_{\rm eff} T_{\rm m}}}{2(1+\sqrt{1+Z_{\rm eff} T_{\rm m}})/Z_{\rm eff}+2T_{\rm m}+T_{\rm h} \sqrt{1+Z_{\rm eff} T_{\rm m}}}
\nonumber \\
&=
\frac{\Delta T \sqrt{1+Z_{\rm eff} T_{\rm m}}(\sqrt{1+Z_{\rm eff} T_{\rm m}}-1)}
{(2T_{\rm m}-T_{\rm h} )\sqrt{1+Z_{\rm eff} T_{\rm m}}+T_{\rm h} (1+Z_{\rm eff} T_{\rm m})}
\nonumber \\
&=\left(\frac{\Delta T}{T_{\rm h}}\right)
\frac{\left(\sqrt{1+Z_{\rm eff} T_{\rm m}}-1 \right)}{ \sqrt{1+Z_{\rm eff} T_{\rm m}}+(T_{\rm c}/T_{\rm h})} ,
\label{eq:etamax}
\end{eqnarray}
where $\Delta T/T_{\rm h}$ is the Carnot efficiency; 
the maximum efficiency approaches the Carnot efficiency by increasing $Z_{\rm eff}$.  
By introducing $\alpha_{\rm eff}$ defined by Eq. (\ref{eq:alphaeff}),  the conventional form of the maximum conversion efficiency given by 
Eq. (\ref{eq:etamax}) is obtained, 
where $Z=\mbox{zT}_{\rm b}$ given by Eq. (\ref{eq:zTbulk}) is redefined by Eq. (\ref{eq:Zeff});  
it is sufficient to study $Z_{\rm eff}$ in place of $Z=\alpha^2/(RK)$ with $R=\rho d/S$ and $K=\kappa S/d$ for films. 
The maximum conversion efficiency is higher for the higher value of $Z_{\rm eff}$. 

The temperature difference at both ends of the material is given by $T(0)-T(d)$ which is related to the ambient temperature difference $\Delta T$ 
by Eq. (\ref{eq:dT}). 
The factor $1+\kappa/(h d)$ in Eq. (\ref{eq:dT}) is absorbed in $\alpha_{\rm eff}$ in Eq. (\ref{eq:alphaeff}). 
Using $\alpha_{\rm eff}$, $Z_{\rm eff}$ is defined and 
the maximum conversion efficiency is a monotonically increasing function of $Z_{\rm eff}$ for any thickness of the thermoelectric material. 
If we use $\alpha$ in stead of $\alpha_{\rm eff}$ in $Z_{\rm eff}$, 
the correlation to the maximum conversion efficiency is lost. 
Therefore, it is essential to define $Z_{\rm eff}$ in terms of $\alpha_{\rm eff}$ 
when the heat transfer coefficient is low for thin film ($h d<\kappa$). 

By taking into account the contact electrical resistivity and the heat transfer coefficient, 
 the effective thermoelectric figure of merit for film conductors [Eq. (\ref{eq:Zeff})] can be expressed as 
 $\mbox{zT}_{\rm eff}(d)=Z_{\rm eff}T_{\rm m}$ 
 and $\mbox{zT}_{\rm eff}(d)$ is explicitly expressed as 
 \begin{eqnarray}
\mbox{zT}_{\rm eff}(d)&=\frac{\alpha_{\rm eff}^2 T_{\rm m}}{\rho_{\rm eff} \kappa_{\rm eff}}, 
\label{eq:ztf0}\\
&= \mbox{zT}_{\rm b} S_{\rm zT} (d), 
\label{eq:ztf}
\end{eqnarray}
where $\mbox{zT}_{\rm b}$ is the figure of merit of bulk given by Eq. (\ref{eq:zTbulk}) and 
the size factor for the effective figure of merit can be expressed as 
\begin{eqnarray}
S_{\rm zT} (d)=\frac{1}{[1+\rho_{\rm c}/(\rho d)][1+\kappa/(hd)]}.
\label{eq:fzT}
\end{eqnarray}
The thickness dependence of $\mbox{zT}_{\rm eff}(d)=Z_{\rm eff}T_{\rm m}$ can be studied using $S_{\rm zT} (d)$;  
the size factor of the effective figure of merit [Eq. (\ref{eq:fzT})] is a monotonically increasing function of the thickness $d$ of the thermoelectric generating layer.

It should be stressed that the conventional figure of merit defined only by the material parameters [$\mbox{zT}_{\rm b}$ in Eq. (\ref{eq:zTbulk})] 
is not a good index of the maximum conversion efficiency for thin film conductors; 
the appropriate index is given by Eq. (\ref{eq:ztf}), where 
the size factor given by Eq. (\ref{eq:fzT}) is taken into account. 
The equivalent expression of Eq. (\ref{eq:fzT}) was defined to capture the interface effects in thermoelectric microrefrigerators. \cite{Sungtaek_00}

We also study the power factor of thin film by maximizing the electrical output power. 
The maximum power is obtained for $y=1$ in Eq. (\ref{eq:power}) and is given by 
\begin{eqnarray}
P_{\rm max}=\frac{(\alpha_{\rm eff} \Delta T)^2}{4R_{\rm eff}}  .
\label{eq:powermax}
\end{eqnarray}
The condition $y=1$ leads to $R_{\rm ext}= R_{\rm eff}$, where $R_{\rm eff}=(\rho +\rho_{\rm c}/d)d/S$ involves 
the contact resistivity. 
In general, the external load resistance should match 
the effective resistance involving both 
the contact resistivity and the intrinsic material resistivity. 
Equation (\ref{eq:powermax})  indicates that the thickness dependence of the maximum electric output power can be 
studied by $\alpha_{\rm eff}^2/R_{\rm eff}$. 
Conventionally,  the bulk power factor is defined by 
\begin{eqnarray}
\mbox{Pf}_{\rm b}=\frac{\alpha^2}{\rho} , 
\label{eq:pfb}
\end{eqnarray}
where $P_{\rm max}$ is multiplied by $d$  to eliminate the trivial thickness dependence originating from $R=\rho d/S$, which remains even for  
$\rho_{\rm c}=0$. 
As a result, the bulk power factor is expressed only by material parameters such as  $\rho$ and $\alpha$. 
Using Eq. (\ref{eq:powermax}), we obtain the power factor of thin film with thickness $d$ as  
$
\mbox{Pf}_{\rm eff}(d)=\alpha_{\rm eff}^2/\rho_{\rm eff}$. 
However, it is more appropriate to define 
an effective power factor using  $\mbox{Pf}_{\rm g}(d)=\alpha_{\rm eff}^2/R_{\rm eff}$ 
to absorb all terms associated with $d$ for studying the thickness dependence. 
We rewrite $\mbox{Pf}_{\rm g}(d)=\mbox{Pf}_{\rm eff}(d)/d$ as 
\begin{eqnarray}
\mbox{Pf}_{\rm g}(d)=\mbox{Pf}_{\rm b}S_{\rm Pw} (d),
\label{pffilm}
\end{eqnarray}
where the size factor for the effective power factor is given by 
\begin{eqnarray}
S_{\rm Pw} (d)=\frac{1}{d[1+\rho_{\rm c}/(\rho d)][1+\kappa/(hd)]^2} . 
\label{eq:fPw}
\end{eqnarray}
The equivalent expression of Eq. (\ref{eq:powermax}) was introduced previously by noticing the temperature difference between the actual device temperature and 
the ambient temperature.  \cite{Mukaida_20}

For a given thickness $d$, both $\rho_{\rm c}<\rho d$ (small enough contact resistance) and $h>\kappa/d$ (high enough heat transfer coefficient) should be satisfied 
to associate the maximum conversion efficiency and the maximum power with the bulk dimensionless thermoelectric figure of merit [Eq. (\ref{eq:zTbulk})] 
and the bulk power factor [Eq. (\ref{eq:pfb})], respectively. 
For the maximum conversion efficiency and the maximum power, 
the effect of finite thickness should be considered either when $d<\rho_{\rm c}/\rho$ or when $d<\kappa/h$ holds. 

Contrary to the size factor of the effective figure of merit [Eq. (\ref{eq:fzT})], which is a monotonically increasing function of the thickness $d$,  
the size factor of the effective power factor [Eq. (\ref{eq:fPw})] has a maximum at the thickness of the thermoelectric generating layer given by  
\begin{eqnarray}
d_{\rm max}=\frac{\kappa/h+\sqrt{\kappa/h\left(\kappa/h+8\rho_{\rm c}/\rho \right)}}{2}. 
\label{eq:maxd}
\end{eqnarray}
When $\kappa/h>8\rho_{\rm c}/\rho$ is satisfied, the above equation is simplified to 
\begin{eqnarray}
d_{\rm max} \approx \kappa/h .
\label{eq:maxda}
\end{eqnarray}
The appearance of the maximum in the output power can be understood in the following way. 
When the film thickness is small, 
the actual temperature difference at the both ends of the film is smaller than the ambient temperature difference as shown in Eq. (\ref{eq:dT}). 
As a result, the Seebeck power is reduced. 
By increasing the film thickness, the actual temperature difference increases and 
the Seebeck power also increases. 
Therefore, the output power increases up to a certain film thickness. 
When the film thickness exceeds $d_{\rm max}$, 
the electric power output decreases with increasing the film thickness because 
the electrical resistance [$R_{\rm eff}=(\rho +\rho_{\rm c}/d)d/S$] increases with $d$. 

\begin{table}[h]
\caption{The physical properties of inorganic materials (Bi$_2$Te$_3$) and organic materials (PEDOT)}
\centering
\begin{tabular}{lrr}
\hline
Transport coefficient & Bi$_2$Te$_3$ & PEDOT\\
\hline \hline
Electrical resistivity $\rho$ [$\Omega$cm]& $1\times 10^{-3} \mbox{ }^{\rm b} $ & $2.8 \times 10^{-2}, 1.2 \times 10^{-3} \mbox{ }^{\rm e}$ \\
Thermal conductivity $\kappa$ [W/(mK)]& $2.0\mbox{ }^{\rm b} $ &$0.1 (0.17), 0.9 (0.94)$ $\mbox{ }^{\rm f}$ \\
Electrical contact resistivity$^{\rm a}$ $\rho_{\rm c}$ [$\mu \Omega$ cm$^2$] & $10^2$ $\mbox{ }^{\rm c}$ & $(1- 20)\times 10^4$ $\mbox{ }^{\rm g}$\\
Heat transfer coefficient$^{\rm a}$ $h$ [W/(m$^2$K)] & $1-100$ $^{\rm d}$& $1-100$ $^{\rm d}$\\
\hline
\end{tabular}
\begin{itemize}
\item[] $^{\rm a}$ Both the electrical contact resistivity and heat transfer coefficient should be defined 
at the interface between the material and the environment. Here, typical values 
are shown without specifying the environment. 
\item[] $^{\rm b}$ \cite{Witting_19}
\item[] $^{\rm c}$ \cite{Khedim_21}
\item[] $^{\rm d}$\cite{Wang_20,Mukaida_20}
\item[] $^{\rm e}$ Through plane value is $2.8 \times 10^{-2}$ and in-plane value is $1.2 \times 10^{-3}$. \cite{Wei_14ACS}
\item[] $^{\rm f}$ Through plane value is 0.1 (0.17) and the in-plane value is 0.9 (0.94). \cite{WEI_16_1,Wei_16} 
(The values in the parenthesis are from ref. \cite{Wei_16}.)
\item[] $^{\rm g}$ The values depend on the combination of PEDOT and the contacted material. \cite{MUKAIDA_17}
\end{itemize}
\label{table:1}
\end{table}
We study the size factor of the effective figure of merit [Eq. (\ref{eq:fzT})] and the  size factor of the effective power factor [Eq. (\ref{eq:fPw})] 
for the benchmark organic/inorganic thermoelectric materials.
Specifically, we use the physical properties of inorganic materials (Bi$_2$Te$_3$) and organic materials (PEDOT) summarized in Table \ref{table:1}. 
The thickness is varied from mm to cm range according to Ref. \cite{Mukaida_20}.
The maxima calculated from Eq. (\ref{eq:maxd}) lie in this length scale by substituting the material properties in Table \ref{table:1}.

\begin{figure}[h]
\includegraphics[width=12cm]{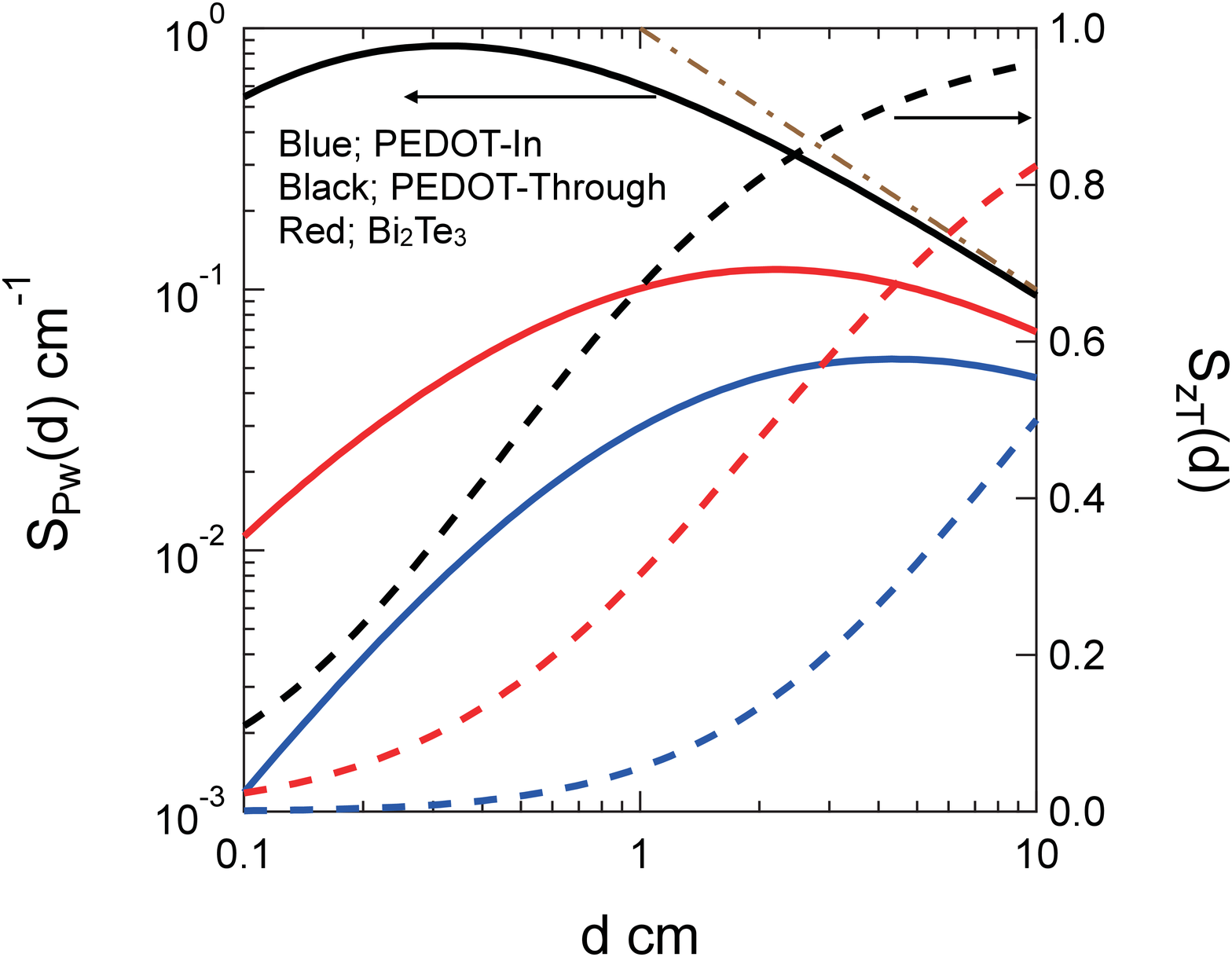}
\caption{(Color online) $S_{\rm Pw} (d)$ (the size factor for the effective power factor) and 
$S_{\rm zT} (d)$ (the size factor for the effective figure of merit) are shown against $d$.
The blue lines indicate PEDOT with the temperature gradient in the in-plane direction, 
the black lines indicate PEDOT with the temperature gradient in the through-plane direction, and 
the red lines indicate Bi$_2$Te$_3$. (PEDOT-Through, Bi$_2$Te$_3$, and PEDOT-In from top to bottom.)
The material parameters are shown in Table \ref{table:1}, where the values without parenthesis, 
$h=100$ [W/(m$^2$K)] for both Bi$_2$Te$_3$ and PEDOT, and $\rho_{\rm c}=1.0 \times 10^4$ $\mu \Omega$ cm$^2$ for PEDOT are used. 
The brown dashed-and-dotted line indicates $S_{\rm Pw} (d)=1/d$ originating from the trivial thickness dependence of $R=\rho d/S$, 
where we assume $\rho_{\rm c}=0$ and take  the limit of $h\rightarrow \infty$. 
$S_{\rm zT} (d)$ in the limit of $d\rightarrow \infty$ is $1$.
}
\label{fig:SpSz}
\end{figure}

As shown in Fig. \ref{fig:SpSz}, $S_{\rm zT} (d)$ (the size factor for the effective figure of merit) monotonically increases by increasing $d$ for 
both  inorganic materials (Bi$_2$Te$_3$) and organic materials (PEDOT). 
By considering anisotropy in intrinsic material electrical resistivity and intrinsic material thermal conductivity of PEDOT, 
we show the case that the temperature gradient is applied in the in-plane direction, 
and the case  that the temperature gradient is applied in the through-plane direction for PEDOT, where  PEDOT would tend to lie down at the surface. 
In the same figure, we also show $S_{\rm Pw} (d)$ (the size factor for the effective power factor). 
$S_{\rm Pw} (d)$ shows a maximum as a function of the film thickness for both  inorganic materials (Bi$_2$Te$_3$) and organic materials (PEDOT). 
Judging from $S_{\rm Pw} (d)$, 
PEDOT with the temperature gradient in the through-plane direction (PEDOT-Through) can be optimized at the thinner film thickness compared to the other materials. 
The optimum value of $d$ for $S_{\rm Pw} (d)$ roughly corresponds to the inflection point for $S_{\rm zT} (d)$, 
indicating that the decrease of $S_{\rm zT} (d)$ by decreasing $d$ mainly occurs around the optimum value of $d$.
The output power can be optimized around $1$ cm for Bi$_2$Te$_3$ and could be optimized below $1$ cm for PEDOT-Through. 
Though the effective figure of merit is a monotonically increasing function of $d$, 
the large part of the figure of merit can be reduced by reducing the thickness from $10$ to $1$ cm for Bi$_2$Te$_3$ and the figure of merit can be largely reduced 
by reducing the thickness around 1 cm for PEDOT-Through. 
By employing the finite element methods, the results similar to Fig. \ref{fig:SpSz} was reported, where 
the maximum efficiency monotonically increases by increasing $d$ and the maximum power shows a maximum as a function of $d$. \cite{Wang_20}

\begin{figure}[h]
\includegraphics[width=8cm]{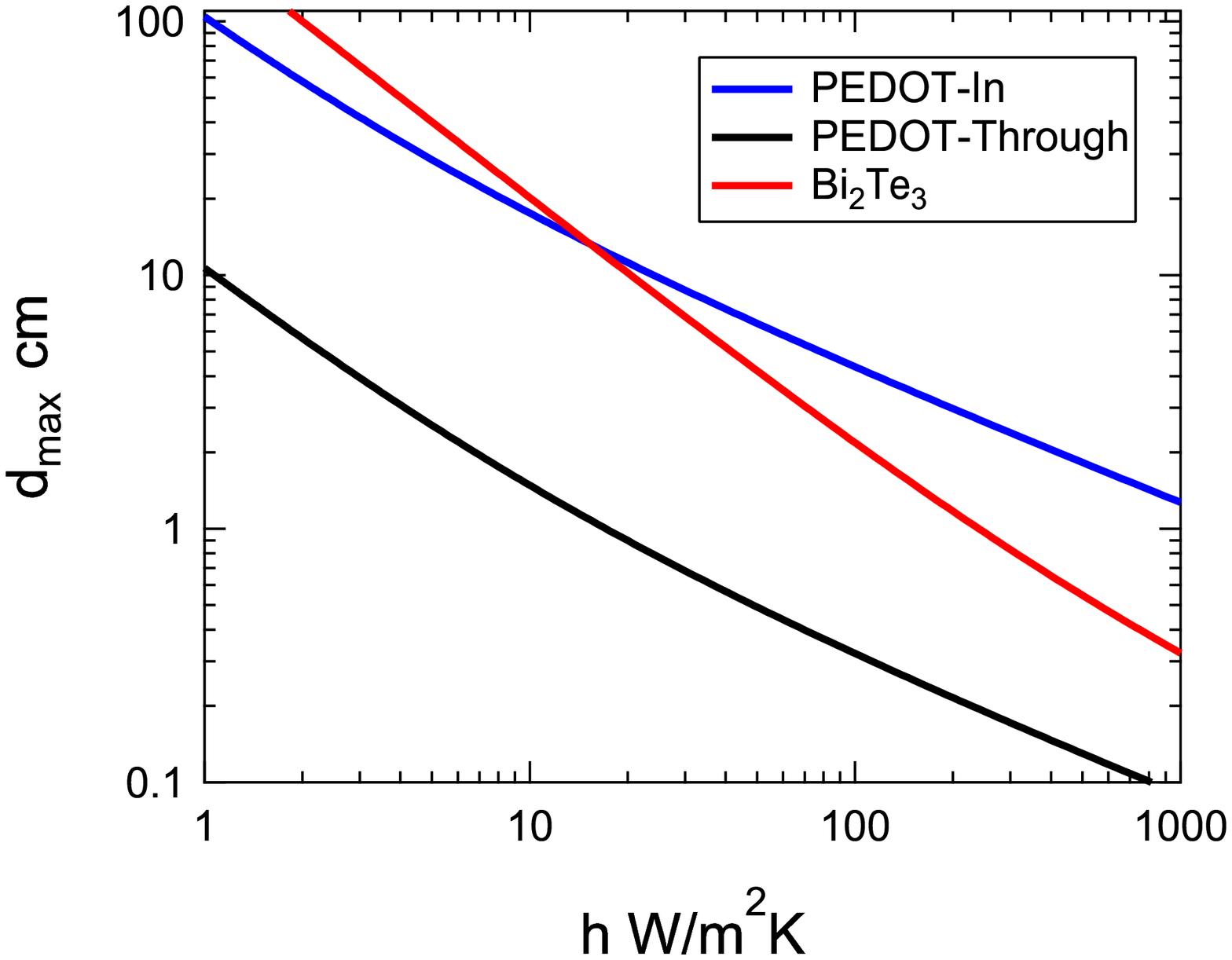}
\caption{(Color online) $d_{\rm max}$ is presented as a function of the heat transfer coefficient ($h$). 
The blue line indicates PEDOT with the temperature gradient in the in-plane direction, 
the black line indicates PEDOT with the temperature gradient in the through-plane direction, and 
the red line indicates Bi$_2$Te$_3$. 
(PEDOT-In, Bi$_2$Te$_3$, and PEDOT-Through from top to bottom at $h=100$W/m$^2$K.)
The material parameters are shown in Table \ref{table:1}, where the values without parenthesis, 
and $\rho_{\rm c}=1.0 \times 10^4$ $\mu \Omega$ cm$^2$ for PEDOT are used. 
}
\label{fig:dmaxvsh}
\end{figure}

In Fig. \ref{fig:dmaxvsh}, we study the film thickness denoted by $d_{\rm max}$ at the maximum in $S_{\rm Pw} (d)$ as a function of the heat transfer coefficient denoted by $h$.
$d_{\rm max}$ decreases by increasing the heat transfer coefficient because 
the temperature gradient could be closer to the imposed temperature gradient 
applied through ambient temperature if the heat transfer is higher. 
In  Fig. \ref{fig:dmaxvsh}, $d_{\rm max}$ of PEDOT-Through is smaller than $d_{\rm max}$ of Bi$_2$Te$_3$, though 
the difference decreases by increasing $h$. 

\begin{figure}[h]
\includegraphics[width=8cm]{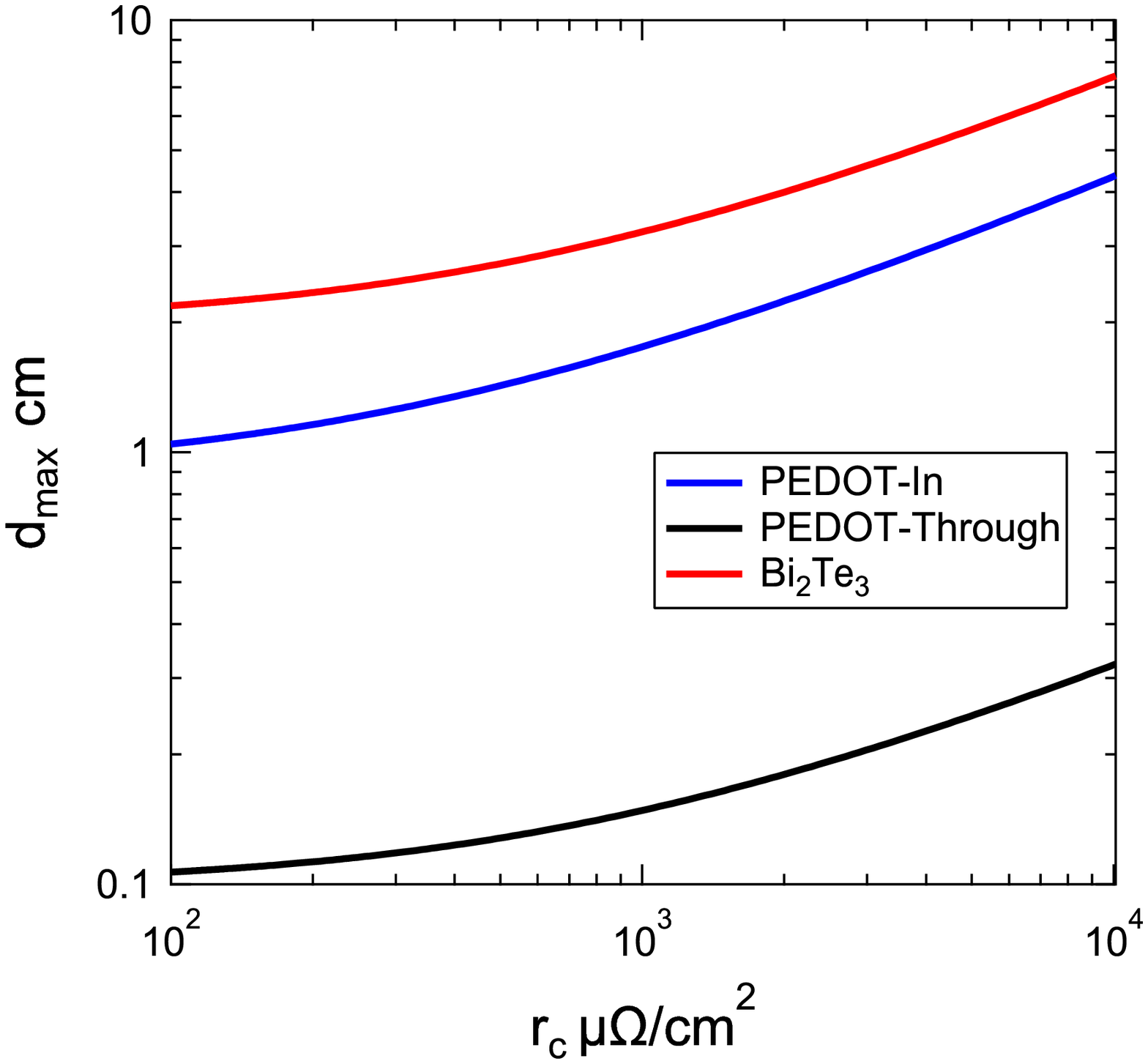}
\caption{(Color online) $d_{\rm max}$ is presented as a function of the contact resistivity ($r_{\rm c}$). 
The blue line indicates PEDOT with the temperature gradient in the in-plane direction, 
the black line indicates PEDOT with the temperature gradient in the through-plane direction, and 
the red line indicates Bi$_2$Te$_3$.  (Bi$_2$Te$_3$, PEDOT-In, and PEDOT-Through from top to bottom.)
The material parameters are shown in Table \ref{table:1}, where the values without parenthesis, and 
$h=100$ [W/(m$^2$K)] are used. 
}
\label{fig:dmaxvsrc}
\end{figure}

In Fig. \ref{fig:dmaxvsrc}, we study the film thickness denoted by $d_{\rm max}$ at the maximum in $S_{\rm Pw} (d)$ as a function of the contact resistivity  denoted by $r_{\rm c}$.
$d_{\rm max}$ increases by increasing $r_{\rm c}$. 
$d_{\rm max}$ of PEDOT-Through is more than an order of magnitude smaller than $d_{\rm max}$ of Bi$_2$Te$_3$. 

In practice, thermoelectric generators may be composed of series connected many semiconductors. 
When the each unit is identical, 
$S_{\rm Pw} (d)$ and $S_{\rm zT} (d)$ of the $N$ repeated units are independent of $N$ as shown below. 
If the electrical resistivity, the thermal conductivity, the electrical contact resistivity, and the heat transfer coefficient of 
the unit are 
denoted by $\rho_{\rm u}$, $\kappa_{\rm u}$, $\rho_{\rm c,u}$, and $h_{\rm u}$, 
those for the electrically connected units in series are expressed as 
$N \rho_{\rm u}$, $N \kappa_{\rm u}$, $N \rho_{\rm c,u}$, and $N h_{\rm u}$.
Because $S_{\rm Pw} (d)$ and $S_{\rm zT} (d)$ of the $N$ repeated units are expressed in terms of  
the ratio such as $\kappa/h$ and $\rho/\rho_{\rm c}$, 
they are independent of $N$. 
However, the dimensionless figure of merit for a single material should be generalized 
if thermoelectric power generator consists of series connected p-type and n-type semiconductors. 
We introduce the effective Seebeck coefficient $\alpha_{\rm eff,p}=\alpha_{\rm p}/[1+\kappa_{\rm p}/(h_{\rm p} d)]$, 
effective electrical resistivity $\rho_{\rm eff,p}=\rho_{\rm p}+\rho_{\rm c,p}/d$, and effective thermal conductivity 
$\kappa_{\rm eff,p}=(1/\kappa_{\rm p}+1/(h_{\rm p}d))^{-1}$ of the p-type semiconductor, 
where the subscript p indicates the property of p-type semiconductor;  
the material properties for the n-type semiconductor are defined similarly. 
When the p-type semiconductor with the area $S_{\rm p}$ and n-type semiconductor with the area $S_{\rm n}$ are 
connected electrically in series and thermally in parallel, while the thickness $d$ is common for both semiconductors, 
$Z_{\rm eff}=(\alpha_{\rm eff}^2 T_{\rm m})/(R_{\rm eff} K_{\rm eff})$ given by Eq. (\ref{eq:Zeff}) still holds, 
where we have $R_{\rm eff} =\rho_{\rm eff,p}d/S_{\rm p} + \rho_{\rm eff,n}d/S_{\rm n}$,  
$K_{\rm eff}=\kappa_{\rm eff,p} S_{\rm p}/d+\kappa_{\rm eff,n} S_{\rm n}/d$ and 
$\alpha_{\rm eff}=\alpha_{\rm eff,p}-\alpha_{\rm eff,n}$; 
the minus sign in front of $\alpha_{\rm eff,n}$ is introduced to take into account the opposite direction of electron flow in n-type semiconductor 
against the direction of hole flow in p-type semiconductor.
If $R_{\rm eff} K_{\rm eff}$ is minimized by optimizing the ration $S_{\rm p}/S_{\rm n}$, 
we find 
$\mbox{zT}_{\rm eff}(d)=\left(\alpha_{\rm eff,p}-\alpha_{\rm eff,n} \right)^2 T_{\rm m}/
[\sqrt{\rho_{\rm eff,p} \kappa_{\rm eff,p}}+\sqrt{\rho_{\rm eff,n} \kappa_{\rm eff,n}}]^2$, 
where 
$S_{\rm p}/S_{\rm n}=\sqrt{\rho_{\rm p}\kappa_{\rm n}/(\rho_{\rm n}\kappa_{\rm p})}$ is obtained by
$d(R_{\rm eff} K_{\rm eff})/dX$ using $X=S_{\rm p}/S_{\rm n}$.
In the limit of $d \rightarrow \infty$, the known result is recovered. \cite{Ioffe}

We studied 
the case that the intrinsic electrical resistivity, thermal conductivity (regardless of  phonon thermal conductivity or electronic thermal conductivity), and Seebeck coefficient are material properties  
independent of the dimensions. 
On this basis, we focused 
to illustrate the genuine contributions of heat transfer coefficient and contact resistivity to the maximum energy conversion efficiency and 
thermoelectric power.

In conclusion, the contact resistivity and heat transfer coefficient at the interface between 
the thermoelectric material and the environment 
influence the energy conversion efficiency and the electric output power if thermoelectric materials are thin and having  
a large surface area.  
The conventional thermoelectric figure of merit and the power factor are not sufficient as a measure of the thin film quality of thermoelectric materials. 
The effective thermoelectric figure of merit of thin film was introduced in such a way that 
the maximum conversion efficiency is a monotonically increasing function of the effective thermoelectric figure of merit regardless of the thickness of the thermoelectric material. 
The effective thermoelectric figure of merit thus defined can be expressed as a product of the conventional thermoelectric figure of merit and the size factor. 
The size factor of the effective thermoelectric figure of merit is shown to be an increasing function of the film thickness. 
Similarly, we introduced the effective power factor and the corresponding size factor; 
we showed the existence of the optimal film thickness. 
 We studied the thickness dependence of the size factor of the effective thermoelectric figure of merit and 
 the size factor of the effective power factor using the physical properties of inorganic materials (Bi$_2$Te$_3$) and organic materials (PEDOT). 
We showed that PEDOT with the temperature gradient in the through-plane direction is advantageous over Bi$_2$Te$_3$ 
as regards to the thinner optimal film thickness.  

\appendix
\section{Derivation of Eq. (\ref{eq:dT})}
\label{Sec:ApA}
The boundary condition at $x=0$ is given by
\begin{eqnarray}
\left. \kappa \frac{\partial T}{\partial x} \right|_{x=0}=h_0\left[T(0)-T_{\rm h}\right],
\label{eq:bc1}
\end{eqnarray}
where the heat flux flowing into the film is considered. 
The boundary condition at $x=d$ is given by 
\begin{eqnarray}
\left. -\kappa \frac{\partial T}{\partial x} \right|_{x=d}=h_d\left[T(d)-T_{\rm c}\right],
\label{eq:bc2}
\end{eqnarray}
where the heat flux flowing out from the film is considered. 
The steady state solution of the heat equation $\nabla^2 T(x)=0$ can be expressed as 
$T(x)=C_1+C_2 x$. 
The two unknown constants $C_1$ and $C_2$ can be determined from the two boundary conditions [Eq. (\ref{eq:bc1}) and Eq. (\ref{eq:bc2})] and we obtain 
\begin{eqnarray}
T(0)&=\frac{T_{\rm h}\left[1+\kappa/(h_d d) \right]+T_{\rm c}\kappa/(h_0 d) }{1+\kappa/(h_d d) +\kappa/(h_0 d) } ,
\label{eq:T0}\\
T(d)&=\frac{T_{\rm c}\kappa/(h_d d)+T_{\rm c} \left[1+\kappa/(h_0 d) \right]}{1+\kappa/(h_d d) +\kappa/(h_0 d) } .
\label{eq:Td}
\end{eqnarray}
we find  Eq. (\ref{eq:dT}) from Eqs. (\ref{eq:T0}) and (\ref{eq:Td}). 

\nocite{*}
\section*{References}
\bibliographystyle{iopart-num}


\end{document}